\documentclass[aps,pre,10pt,twocolumn,noshowpacs,amsfonts,amssymb,amsmath,longbibliography,superscriptaddress,notitlepage]{revtex4-1}
\usepackage[utf8]{inputenc}
\usepackage{bm}

\usepackage{natbib}
\usepackage[hidelinks]{hyperref}
\usepackage{graphicx}
\usepackage[dvipsnames]{xcolor}

\usepackage{array}
\usepackage{algorithm}
\usepackage{algpseudocode}
\usepackage{comment}

\newcommand{\vecp}{\mathbf{p}}

\newcommand{\zhat}{\mathbf{\hat{z}}}
\newcommand{\smat}{\mathbf S}

\begin{document}

\title{Sensitivity analysis of epidemic forecasting and spreading on networks with probability generating functions}

\author{Mariah C. \surname{Boudreau}}
\affiliation{Vermont Complex Systems Institute, University of Vermont, Burlington VT}
\affiliation{Department of Mathematics \& Statistics, University of Vermont, Burlington VT}
\author{William H. W. \surname{Thompson}}
\affiliation{Vermont Complex Systems Institute, University of Vermont, Burlington VT}
\author{Christopher M. \surname{Danforth}}
\affiliation{Vermont Complex Systems Institute, University of Vermont, Burlington VT}
\affiliation{Department of Mathematics \& Statistics, University of Vermont, Burlington VT}
\author{Jean-Gabriel \surname{Young}}
\affiliation{Vermont Complex Systems Institute, University of Vermont, Burlington VT}
\affiliation{Department of Mathematics \& Statistics, University of Vermont, Burlington VT}
\author{Laurent \surname{H\'ebert-Dufresne}}
\affiliation{Vermont Complex Systems Institute, University of Vermont, Burlington VT}
\affiliation{Department of Computer Science, University of Vermont, Burlington VT}

\date{\today}

\begin{abstract}
Epidemic forecasting tools embrace the stochasticity and heterogeneity of disease spread to predict the growth and size of outbreaks. Conceptually, stochasticity and heterogeneity are often modeled as branching processes or as percolation on contact networks. Mathematically, probability generating functions provide a flexible and efficient tool to describe these models and quickly produce forecasts. While their predictions are probabilistic---i.e., distributions of outcome---they depend deterministically on the input distribution of transmission statistics and/or contact structure. Since these inputs can be noisy data or models of high dimension, traditional sensitivity analyses are computationally prohibitive and are therefore rarely used. Here, we use statistical condition estimation to measure the sensitivity of stochastic polynomials representing noisy generating functions. In doing so, we can separate the stochasticity of their forecasts from potential noise in their input. For standard epidemic models, we find that predictions are most sensitive at the critical epidemic threshold (basic reproduction number $R_0 = 1$) only if the transmission is sufficiently homogeneous (dispersion parameter $k > 0.3$). Surprisingly, in heterogeneous systems ($k \leq 0.3$), the sensitivity is highest for values of $R_{0} > 1$. We expect our methods will improve the transparency and applicability of the growing utility of probability generating functions as epidemic forecasting tools.
\end{abstract} 
\maketitle

\section{Introduction}
Epidemic dynamics are stochastic, heterogeneous, noisy, and, at times, very sensitive to small changes in pathogen properties or human behavior.
Our epidemic forecasts are similar and require careful calibration to provide useful predictions.
When epidemic forecasters are asked if another epidemic might occur, a typical response is ``not if, but when.'' 
As a result, the community largely embraces probabilistic forecasts that model the probability that a large epidemic emerges when a pathogen is introduced into a new population.
These forecasts clash with traditional modeling approaches, as they require large simulations or high-dimensional mathematical tools, and consequently, less is known about the sensitivity of their predictions.

Classic compartmental models, such as Kermack and McKendrick's SIR model \cite{kermack1927contribution, kermack1932contributions, kermack1933contributions}, represent deterministic dynamics, eliminating the possibility of stochastic extinctions. 
These deterministic frameworks do not capture epidemics that die out early; instead, outbreaks always occur if transmission parameters exceed a certain threshold, and never do under that threshold. 
To account for stochastic extinction, modelers move to stochastic models, such as branching processes. These types of models, in particular those whose analysis uses probability generating functions (PGFs), measure not ``if'' an epidemic can occur but ``what are the chances of an outbreak of a given size,'' encapsulating the randomness of everyday interactions, extinctions, and extreme events. 
In this work, we use PGFs to estimate the survival probability of an epidemic, which we will refer to as the epidemic probability (or its complement, the extinction probability).

In these stochastic models, we often assume that input parameters are exact observations and lack any noise. We then let uncertainty in our forecast be only a reflection of the stochasticity of the dynamics. The effects of noise in the observed inputs, if any, are intertwined with the probabilistic nature of the forecasts and therefore hard to estimate. Here, we quantify the sensitivity of these models and extend forecasts about ``the chances of an outbreak of a given size'' to include ``their \emph{sensitivity to noise}''. While sensitivity analyses are not novel in deterministic settings like epidemiological compartmental models \cite{lu2023global, gilbert2014probabilistic, powell2005sensitivity}, we present a sensitivity analysis of epidemiological applications of PGFs to provide intuition about the reliability of stochastic and probabilistic forecasts.

\begin{figure*}[t]
    \centering
    \includegraphics[width=\textwidth]{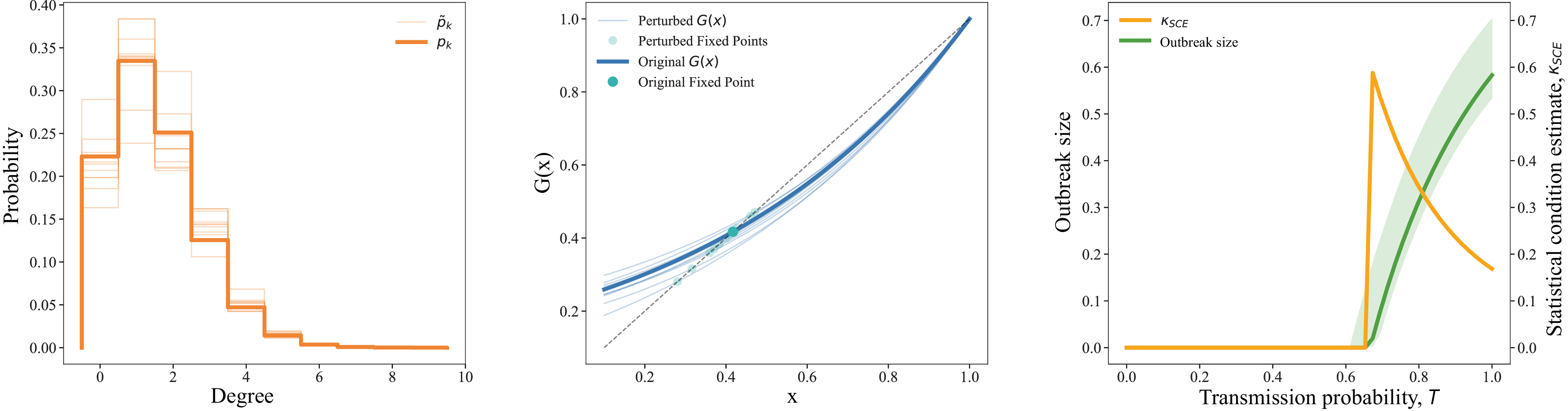}
    \caption[Conceptual framework]{\textbf{Conceptual framework:} The first step of our sensitivity analysis is to perturb the probability distribution generated by the PGF, $G(x)$. Second, we solve for the extinction probability, $u$, for the unperturbed and perturbed PGFs. In branching processes, this probability is given by the fixed point between $[0,1]$ of the self-consistent equation $G(x) = x$. Finally, we compute the statistical condition estimate, $\kappa_{\mathrm{SCE}}$, from the computed fixed points, identifying the distribution parameters and their corresponding probability of survival that are the most sensitive to noise. The shaded region in the far right panel shows a 90\% confidence interval for the relative outbreak size for a percolation example.}
    \label{fig:conceptual_schematic}
\end{figure*}

Sensitivity analyzes help scientists evaluate the limitations and scope of their models. As new methodologies emerge, modelers must explore input parameter spaces to identify how small changes affect their projected outcomes. For models of disease spread, perturbations to input values can affect key metrics such as case counts, mortality rates, or the infected proportion of the population. This is especially important given that epidemiological parameters and the size of the vulnerable population are hard to estimate from early data~\cite{case2023accurately}. Examining these effects helps modelers pinpoint which parameters and, in turn, which particular diseases, are the most sensitive to noise.

We conduct a global sensitivity analysis to evaluate how output variability is apportioned to all sources of input uncertainty over the entire input space \cite{saltelli2008global, wu2013sensitivity}. 
PGF models of disease spread are usually parametrized by probability distribution over the number of infections caused by individuals, encoded as the coefficients of a polynomial \cite{wilf2005generatingfunctionology}. Standard mathematical manipulations briefly detailed in Sec.~\ref{sec:methods} can then be used to deduce the extinction probability and the epidemic probability implied by these case distributions. With a probability distribution as our input, we use a variation of the elementary effect, also known as the Morris method~\cite{morris1991factorial}, to evaluate the sensitivity of our epidemic probability forecasts to variations in model parameters. 
This method produces a \emph{statistical condition estimation} (SCE) of a condition number~\cite{laub2008statistical}, denoted as $\kappa_{\mathrm{SCE}}$.
A visual and conceptual representation of the SCE process is shown in Fig.~\ref{fig:conceptual_schematic}. 

In this paper, we calculate $\kappa_{\mathrm{SCE}}$, for two common applications of PGFs forecasts: Branching processes and percolation on random graphs. Section \ref{sec:pgfFormalism1} describes the PGF method for the former case, while Sec. \ref{sec:pgfFormalism2} details the latter. Section \ref{sec:laub-xia} provides background on condition numbers and explains Laub \emph{et al.}'s algorithm for SCE \cite{laub2008statistical}. The results for each case study follow in Sec.~\ref{sec:negative_binom_lloyd-smith} and \ref{sec:ERsims}. Finally, in Sec.~\ref{sec:discussion}, we discuss how sensitivity patterns differ between homogeneous and heterogeneous systems, with the latter showing maxima of sensitivity far from their critical point.

% ~~~~~~~~~~~~~~~~~~~~~~~~~~~~~~~~~~~~~~~~~~~~~~~~~~~~~~~~~~~~~~~~~~~~~
\section{Methods}
\label{sec:methods}
% ~~~~~~~~~~~~~~~~~~~~~~~~~~~~~~~~~~~~~~~~~~~~~~~~~~~~~~~~~~~~~~~~~~~~~

We review two flavors of PGFs applied to epidemic dynamics and forecasting, describing branching processes or percolation dynamics on random networks. PGFs compactly encode a discrete probability distribution as the coefficients of a formal power series~\cite{wilf2005generatingfunctionology}. In both of our applications, we consider $G(x)$ as the PGF of the \emph{offspring distribution} $\lbrace p_\ell \rbrace$, denoting the probability that an infected individual infects $\ell$ additional individuals in the next generation. We thus write
\begin{equation} \label{eq:G0}
    G(x) = \sum^{\infty}_{\ell=0} p_{\ell}x^{\ell} \; .
\end{equation}
PGFs allow for the easy calculation of many properties of distributions \cite{wilf2005generatingfunctionology}. For example, the average number of offspring (i.e., the basic reproduction number $R_0$) can be calculated by
\begin{equation}
    R_0 \equiv \langle \ell \rangle = \sum \ell p_\ell = \frac{d}{dx}G(x)\bigg\vert_{x=1} = G'(1)\; .
\end{equation}
In what follows, we use similar calculations to derive important quantities of interest for epidemic forecasting.

% ~~~~~~~~~~~~~~~~~~~~~~~~~~~~~~~~~~~~~~~~~~~~~~~~~~~~~~~~~~~~~~~~~~~~~
\subsection{Probability generating functions for branching processes}
\label{sec:pgfFormalism1}
% ~~~~~~~~~~~~~~~~~~~~~~~~~~~~~~~~~~~~~~~~~~~~~~~~~~~~~~~~~~~~~~~~~~~~~

Branching processes are fully defined by their offspring distribution and, therefore, by their PGF, $G(x)$. In the context of epidemic forecasting, each offspring represents a secondary infection. Starting from the first infected individual, patient zero, we are interested in using $G(x)$ to calculate the probability of seeing a stochastic extinction rather than a large macroscopic epidemic that does not end.

We can calculate the \emph{extinction probability} for the branching process, $u_b$, with a self-consistent argument. For an outbreak starting at patient zero to be of finite size, every secondary infection of patient zero must also create chains of transmission of finite sizes. With probability $p_0$, patient zero does not create any secondary infections, and the outbreak is certainly finite. With probability $p_1$, patient zero creates one secondary infection, which must lead to a finite outbreak with probability $u_b$. With probability $p_2$, patient zero creates two secondary infections that \textit{both} lead to a finite outbreak with probability $u_b^2$. And so on and so forth, until we realize that
\begin{equation}
    \label{eq:selfConsistent}
    u_b = \sum_{\ell=0}^\infty p_\ell u_b^\ell = G(u_b) \; ,
\end{equation}
which gives us a self-consistent equation for $u_b$.
We also define the epidemic probability as the complement $1-u_b$. 

Note, Eq.~\eqref{eq:selfConsistent} only admits a non-trivial solution $u_b\in [0,1]$ if the branching process is above its epidemic threshold. In a branching process, this means having an average number of offspring or secondary infections greater than or equal to 1 (i.e., only if $R_0 = G'(1) \geq 1$).

In epidemic forecasting, it is typical to assume that $G(x)$ generates a negative binomial of secondary infections \cite{lloyd2005superspreading,hebert2020beyond}. This particular distribution is shaped by two parameters: Its average, which corresponds to the classic basic reproduction number $R_0$, and its dispersion parameter $k>0$. The dispersion parameter $k$ allows us to shape the heterogeneity of the distribution around its average $R_0$. Counterintuitively, a high dispersion parameter $k$ represents homogeneous distributions (a Poisson distribution in the limit $k\rightarrow\infty$), while a dispersion parameter $k$ close to 0 represents overdispersed distributions with rare but important superspreading events. 
In this parametrization, the probability mass function of the negative binomial is equal to 
\begin{equation}
    \label{eq:nbinom_dist}
    p_{\ell} =  \frac{\Gamma(\ell + k)}{\ell!\ \Gamma(k)} \Big(\frac{R_{0}}{R_{0} + k}\Big)^{\ell} \Big(\frac{k}{R_{0} + k}\Big)^{k}. 
\end{equation}
where $\Gamma(x)$ is the gamma function.
Figure~\ref{fig:negative_binom_roots_outbreaks} below shows forecasts for the probability of an epidemic using $R_{0} \in [0.8, 4]$ and $k \in [0.01, 10]$. 

Importantly, we will want to evaluate the sensitivity of the solution of Eq.~\eqref{eq:selfConsistent} not to the parametrization of $G(x)$ in terms of $R_0$ and $k$, but in terms of the true distribution $\lbrace p_\ell\rbrace$ of the dynamics. This might seem like a minor point, but it is essential for the generality of our approach.

% ~~~~~~~~~~~~~~~~~~~~~~~~~~~~~~~~~~~~~~~~~~~~~~~~~~~~~~~~~~~~~~~~~~~~~
\subsubsection*{Assumptions}
\label{sec:assumptions_NB_pgf}
% ~~~~~~~~~~~~~~~~~~~~~~~~~~~~~~~~~~~~~~~~~~~~~~~~~~~~~~~~~~~~~~~~~~~~~
This framework relies on several assumptions that are critical to keep in mind.
The most important is that Eq.~\eqref{eq:selfConsistent} assumes an infinite population, which consequently means a large-scale epidemic never ends. Additionally, the PGF $G(x)$ does not depend on time, resulting in a constant offspring distribution~\cite{allen2022predicting}. These assumptions collectively imply that the population does not react to the epidemic, there is no depletion of susceptible individuals (given the infinite population), and the pathogen remains genetically stable without mutation.
These assumptions are made to ease the presentation of results, but in future work, the sensitivity analysis techniques presented here could be easily applied to a situation where these assumptions are relaxed.

% ~~~~~~~~~~~~~~~~~~~~~~~~~~~~~~~~~~~~~~~~~~~~~~~~~~~~~~~~~~~~~~~~~~~~~
\subsection{Probability generating functions for percolation on contact networks}
\label{sec:pgfFormalism2}
% ~~~~~~~~~~~~~~~~~~~~~~~~~~~~~~~~~~~~~~~~~~~~~~~~~~~~~~~~~~~~~~~~~~~~~

Percolation on random contact networks is a common way of capturing spreading dynamics over some network structure \cite{newman2002spread, meyers2007contact}. In this context, we consider different PGFs that represent both network structure and the stochastic dynamics of transmission supported by the network. 

Let us first consider our patient zero, a node of the network selected uniformly at random. Its number of contacts $j$ will follow the \emph{degree distribution} $\lbrace p_j \rbrace$ of the network \cite{newman2001random}, generated by
\begin{equation}
    G_0(x) = \sum^{\infty}_{j} p_j x^j \; .
\end{equation}
From there, the neighbors of patient zero are nodes reached by following random edges. Through this process, it is $j$ times more likely to reach a node of degree $j$ than a node of degree one (and therefore obviously impossible to reach a node of degree zero). Patient zero's number of other neighbors therefore follows the \emph{excess degree distribution} of the network \cite{newman2001random} and is generated by
\begin{equation} \label{eq:G1}
G_{1}(x) = \frac{\sum^{\infty}_{j=0} jp_jx^{j-1}}{\sum^{\infty}_{j=0}jp_j} =  
    \frac{1}{\langle j \rangle}G'_{0}(x)\;,
\end{equation}
which is normalized by the average degree, given by
\begin{equation}\label{eq: G0derivative}
     \langle j \rangle = \sum^{\infty}_{j=0} j p_{j} =G'_{0}(1) \; .
\end{equation}

We assume that every contact between an infected individual and their neighbors will lead to transmission with probability $T$. This is a simple Bernoulli trial, with PGF $B(x) = (1-T) + Tx$ where $x=1$ encodes the fact that a transmission event happened. The probability mass function of secondary infection is then determined by the combination of the two stochastic processes---a random number of neighbors, some subset of which get infected.
This distribution is generated by $G(x) = G_1(B(x))$.

This spreading process on a random network is analogous to a branching process, and the probability of extinction after an infection is given by
\begin{equation}
        \label{eq:self_consistent_percolation}
        u = G(u) \equiv G_1((1-T)+Tu)\; .
\end{equation}
However, the overall probability of extinction is defined at patient zero. Having solved for $u$, we write the extinction probability for the percolation process as
\begin{equation}
    u_p = G_0((1-T) + Tu) \; .
\end{equation}

Assuming that the contact network is undirected, there is now a useful equivalence between the probability of a macroscopic epidemic ($1-u_p$) and its eventual size ($S$). Since nodes involved in the largest transmission cluster can either start or be reached by the transmission process, the probability of patient zero being in this cluster is equivalent to the relative size of the cluster. The relative final size of a macroscopic epidemic is equal to the epidemic probability, 
\begin{equation}
    \label{eq:rel_pandemic_size}
    S = 1-u_p \;.
\end{equation}
Hence, solving for $u$ in Eq.~\eqref{eq:self_consistent_percolation} tells us everything there is to know about the process.

Since PGFs are always normalized, $u =1$ will always be a fixed point of Eq.~\eqref{eq:self_consistent_percolation} and in this case, we have $u_p=1$ and $S=1-u_p=0$ since $G_0(1)=1$. There is a second non-trivial solution $u \in [0,1)$ when we exceed the epidemic threshold $R_0 = 1$. Since $R_0$ is defined as the average number of secondary cases, we can write $R_0 = TG_1'(1)$, which gives us the \emph{critical transmission probability},
\begin{equation}
\label{eq:critcal_transition}
    T_c = \frac{1}{G'_1(1)} \; ,
\end{equation}
i.e., the value $T_c$ of $T$ such that $R_0=1$.
This relation allows us to calculate the critical transitions in the rest of the paper. 

The natural question then is to measure how sensitive our forecasts for the probability and size of an epidemic are to our parameterization of network structure, $\lbrace p_j \rbrace$, and disease transmission, $T$.

% ~~~~~~~~~~~~~~~~~~~~~~~~~~~~~~~~~~~~~~~~~~~~~~~~~~~~~~~~~~~~~~~~~~~~~
\subsubsection*{Assumptions}
\label{sec:assumptions_ER_pgf}
% ~~~~~~~~~~~~~~~~~~~~~~~~~~~~~~~~~~~~~~~~~~~~~~~~~~~~~~~~~~~~~~~~~~~~~

One additional assumption to consider for this case study is that the model is only exact on treelike networks: no downstream path can connect two neighbors of a given node since Eq.~(\ref{eq:self_consistent_percolation}) assumes that the secondary cases are independent of one another.

% ~~~~~~~~~~~~~~~~~~~~~~~~~~~~~~~~~~~~~~~~~~~~~~~~~~~~~~~~~~~~~~~~~~~~~
\subsection{Statistical condition estimate of polynomial fixed points}
\label{sec:laub-xia}
% ~~~~~~~~~~~~~~~~~~~~~~~~~~~~~~~~~~~~~~~~~~~~~~~~~~~~~~~~~~~~~~~~~~~~~
We use statistical condition estimation to quantify the sensitivity of extinction probability to noisy data.
In our case studies, this SCE captures the sensitivity of the extinction probability $u$ to perturbations in the distribution $\mathbf{p}$ that parameterizes the PGF models. We encode the probability mass function in a vector $\mathbf{p}=[p_0, p_1, p_2,\hdots, p_m]^\top$, and we denote the function that maps $\mathbf{p}$ to a corresponding extinction probability $u$ as $f(\mathbf{p})$. Note that $f(\mathbf{p})$ is neither the PGF $G(x)$ itself nor the \textit{condition} $G(u) - u = 0$; it is the function that solves for $u$ (see Appendix \ref{sec:solving_self_consitent_equations}).

In the case of the branching process, $u = u_b$ is the self-consistent solution to Eq.~\eqref{eq:selfConsistent}, and $\mathbf{p}$ represents the offspring distribution. In the case of percolation, the extinction probability is given by Eq.~\eqref{eq:self_consistent_percolation}, and $\vecp$ represents the degree distribution of the contact network. In both cases, $m$ is a reasonable upper bound on the number of offspring or the degree of nodes.

The simplest notion of a function's sensitivity is the \emph{absolute condition number}, the magnitude of the gradient at a point, $\| \nabla f(\mathbf{p})\|_2$. It provides a natural measure of the sensitivity of a function to perturbations of the coefficients $\mathbf{p}$. If the condition number of a function is small (large), a small (large) change to the input will result in a correspondingly small (large) change in the output. 

However, this absolute condition number provides a relatively restrictive notion of sensitivity. First, the calculation of the gradient can be difficult or computationally intractable if $f(\mathbf{p})$ either lacks a closed-form expression or if its domain is high-dimensional.  Second, the condition number only considers the sensitivity of functions to isotropic perturbations of constant magnitude, which are not always applicable when first-hand knowledge of the system is available to practitioners.
 
We use the SCE to calculate how the function changes under arbitrary random perturbations to the coefficients of the infection process, $\mathbf{p}$.
We will (ironically) focus on isotopic, uncorrelated perturbations, though we note that the SCE can easily accommodate more general, structured perturbations.
We do this for the sake of the exposition, and because the SCE turns out to be related to the condition number.
Indeed, the SCE is in fact an unbiased estimator of the relative condition number if perturbations follow this specific distribution~\cite{kenney1994small}; see our detailed exposition of this connection in Appendix \ref{sec:cond_num_appendix}.

Our method to calculate the statistical condition estimate is thus as follows \cite{laub2008statistical}:
\begin{enumerate}
    \item[i] Randomly perturb the input $\mathbf{p}$ to $\tilde{\mathbf{p}}$. The size of the perturbation is proportional to a small constant $\delta$: $\|\mathbf{p} -\tilde{\mathbf{p}}\|_2 \propto \delta \ll 1$.
  \item[ii] Calculate the respective extinction probabilities $u(\mathbf{p})$ and $\tilde{u}(\tilde{\mathbf{p}})$.
  \item[iii] Compute the change in the output due to perturbation, normalized by the magnitude of the unperturbed output, 
  \begin{equation*}
      \nu = \frac{\left| u(\bf{p}) - \tilde{u}(\tilde{\bf{p}}) \right|}{\delta |u(\bf{p})| }.
  \end{equation*}
  \item[iv] Repeat $r$ time and use the sampled values $\nu^{(1)}, ...,\nu^{(r)}$ to calculate the relative statistical condition estimate,
    \begin{equation}
         \kappa_{\mathrm{SCE}} \propto \frac{1}{r} \sum_{i=1}^r \nu^{(i)}.
    \end{equation}
\end{enumerate}

It is known that the bounds obtained from $\kappa_{\mathrm{SCE}}$ on the true relative condition number are dramatically tightened if the perturbations are sampled orthogonally and the equation for $\kappa_{\mathrm{SCE}}$ is updated accordingly \cite{kenney1994small}. 
Hence, to calculate our final estimate, we actually generate perturbations as a random orthogonal set of $r$ unit vectors $\{\bf{z}^{(i)}\}$, obtained by calculating the QR decomposition of an $m\times r$ matrix of i.i.d. $N(0,1)$ draws; c.f. Appendix~\ref{sec:rel_cond_num_appendix} for details.
We then ensure the probability distribution is preserved by letting 
\begin{equation}
    \tilde{p}_\ell  = \frac{p'_\ell}{\sum p'_\ell} , \quad p'_\ell = p_\ell(1 + \delta \mathbf{z}_i^{(j)}).
\end{equation}
When the perturbations are sampled so, $\kappa_{\mathrm{SCE}}$ becomes proportional to the 2-norm of the sampled $\nu^{(i)}({\bf{z}})$. 
The constant of proportionality is a ratio of Wallis factors $\omega_n$. 
For any integer $n$, this factor is defined as,
\begin{equation}
    \omega_n = \frac{(n-2)!!}{(n-1)!!}\;,
    \label{eq:wallis_factor_def}
\end{equation}
where $n!!$ is a skip factorial, equal to  $n!! = n \cdot (n-2) \cdot (n-4) \hdots 1$ for odd $n$ and $n!! = n \cdot (n-2) \hdots 2$ for even $n$. 
Our final estimate is thus
\begin{equation}
     \kappa_{\mathrm{SCE}} = \frac{\omega_r}{\omega_m} \sqrt{|\nu^{(1)}|^2 + \hdots + |\nu^{(r)}|^2}\;.
    \label{eq:kappa_sce}
\end{equation}
A condensed SCE algorithm can be found in Appendix~\ref{sec:algorithm}.

% ~~~~~~~~~~~~~~~~~~~~~~~~~~~~~~~~~~~~~~~~~~~~~~
\section{Results}
\label{sec:results}
% ~~~~~~~~~~~~~~~~~~~~~~~~~~~~~~~~~~~~~~~~~~~~~~
We present two case studies inspired by the work of Lloyd-Smith \emph{et al.} \cite{lloyd2005superspreading} and Meyers \cite{meyers2007contact} among others. Lloyd-Smith \emph{et al.} show the importance of individual variation on the distribution of secondary cases. This individual variation can be characterized by a negative binomial offspring distribution encapsulating the individual variation through the dispersion parameter, $k$, which can produce high variance (low $k$) around the average of the distribution ($R_0$) \cite{lloyd2005superspreading, hebert2020beyond}. Meyers models population-level disease spread with bond percolation on heterogeneous random contact networks. Even with a fixed transmission probability for the bond percolation, using fat-tailed degree distributions to specify degrees in the contact network can drive heterogeneity in the spreading process \cite{meyers2007contact, newman2002spread}. We use two random networks and their associated degree distributions in this analysis. The first is an Erd\H{o}s--R\'enyi (ER) random network characterized by a Poisson degree distribution, and the second is a power-law network characterized by a power-law degree distribution.  In both case studies, we test the effect of added error or noise on the input probability distribution.

% ~~~~~~~~~~~~~~~~~~~~~~~~~~~~~~~~~~~~~~~~~~~~~~~~~~~~~~~~~~~~~~~~~~~~~
\subsection{Case Study: Negative binomial branching processes}
\label{sec:negative_binom_lloyd-smith}
% ~~~~~~~~~~~~~~~~~~~~~~~~~~~~~~~~~~~~~~~~~~~~~~~~~~~~~~~~~~~~~~~~~~~~~

\begin{figure*}[th]
    \centering
    \includegraphics[width=\textwidth]{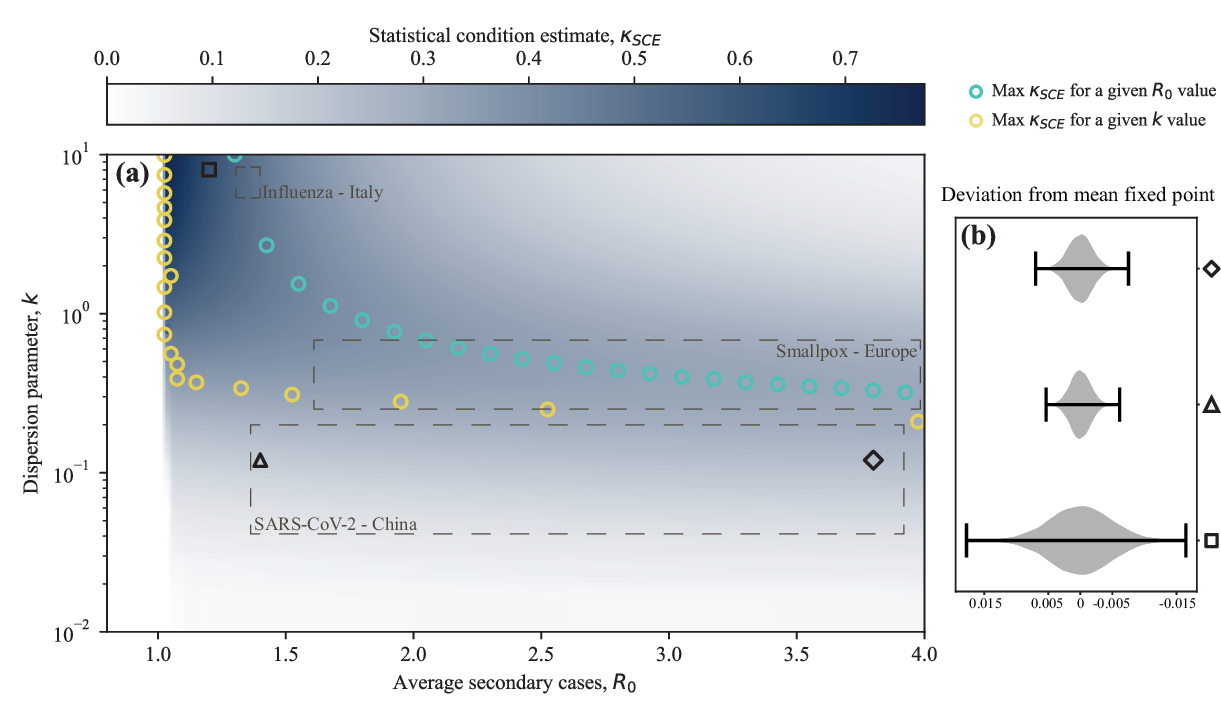}
    \caption[Statistical condition estimates for negative binomial branching processes]{\textbf{Statistical condition estimates for negative binomial branching processes:} \textbf{(a)} The largest $\kappa_{\mathrm{SCE}}$, meaning the most sensitive values, correspond to the ranges $k >  0.3$, and $R_0 \approx 1$ . The maximum $\kappa_{\mathrm{SCE}}$ for a given $R_{0}$ value is shown with the light blue circle markers, which follow a ridge of larger $\kappa_{\mathrm{SCE}}$ from the largest $k$ to $k \approx 0.3$. On the other hand, the maximum $\kappa_{\mathrm{SCE}}$ for a given $k$ value is shown with yellow circle markers. Surprisingly, these markers diverge away from $R_{0} \approx 1$ when $k \leq 0.3$. The black square, triangle, and diamond markers highlight parameterizations studied in panel (b), $(R_{0} = 1.2, k = 8.05)$, $(R_{0} = 1.4, k = 0.12)$, and $(R_{0} = 3.8, k = 0.12)$, respectively. Parameter subspaces, denoted with dashed boxes, depict the likely values of of $R_{0}$ and $k$ for three recent disease outbreaks: the 2020 SARS-CoV-2 outbreak in China, the 1958-1973 Smallpox outbreak in Europe, and the 2009 Influenza outbreak in Italy (see Table \ref{tab:diseases} for details). \textbf{(b)} Violin plots  depicting the deviation from the mean fixed point of 1000 perturbed PGFs with $\delta = 2^{-2}$, for the three parameter pairs highlighted by black markers in panel (a). These violin plots validate the magnitudes of the $\kappa_{\mathrm{SCE}}$ shown in the heat map.}
    \label{fig:negative_binom_cond}
\end{figure*}

\begin{figure}[t!]
    \centering
    \includegraphics[width=\linewidth]{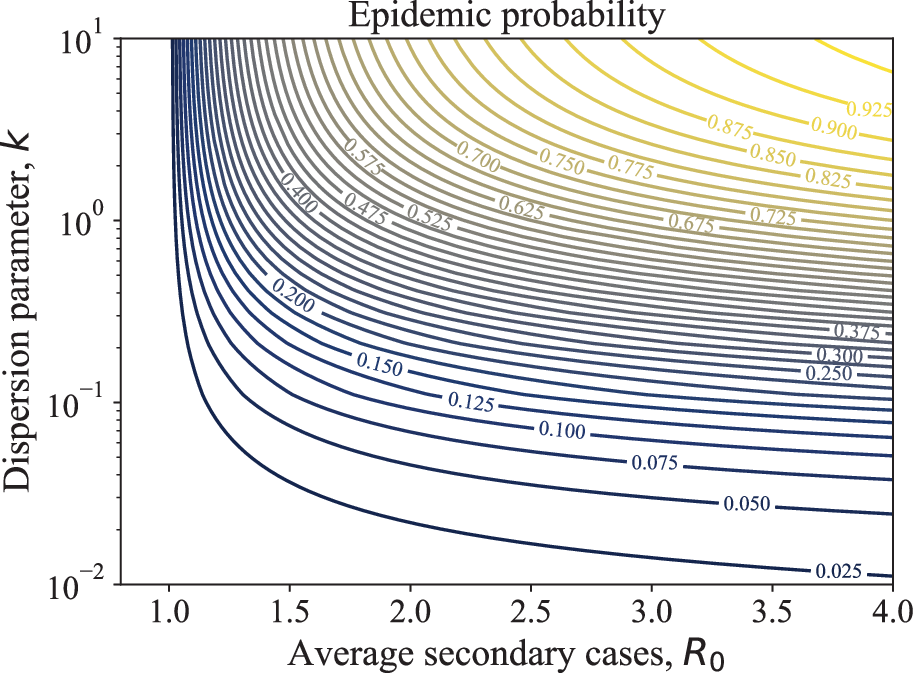}
    \caption[Negative binomial offspring distribution epidemic probability]{\textbf{Negative binomial offspring distribution epidemic probability:}  
    The contour lines show the epidemic probability for a branching process with a negative binomial offspring distribution, parameterized with the average number of secondary cases, $R_{0}$ and a dispersion parameter, $k$. A smaller $k$ defines a more heterogeneous offspring distribution with larger variance. On the other hand, a larger $k$ defines a more homogeneous offspring distribution with smaller variance.
    }
    \label{fig:negative_binom_roots_outbreaks}
\end{figure}

As stated in Sec. \ref{sec:pgfFormalism1}, a negative binomial distribution commonly defines a distribution of secondary cases for public health applications. For the present paper, we limit the range of parameter values to $R_{0} \in [0.8,4]$ and $k \in [0.01, 10]$. This parameter space encapsulates both homogeneous and heterogeneous processes, remembering that homogeneous spread (high $k$) has less variability in the number of secondary cases and heterogeneous spread (low $k$) has more. This leads to a wide range of dynamical outcomes, as shown in Fig.~\ref{fig:negative_binom_roots_outbreaks}. For each pair of parameters, we calculate the $\kappa_{\mathrm{SCE}}$ from 1,000 perturbed PGFs with perturbations using $\delta = 2^{-16}$. To handle the calculation numerically, we truncate the support of the PGF, meaning the number of secondary cases ranges from 0 to 1000. The results are shown in Fig.~\ref{fig:negative_binom_cond}.

Our results show that the sensitivity of homogeneous and heterogeneous processes vary in fundamentally different ways with respect to the average number of secondary infections $R_0$.
In panel (a) of Fig. \ref{fig:negative_binom_cond}, the darkest shades represent the largest $\kappa_{\mathrm{SCE}}$, while the lightest shades represent the smallest. As a reminder, a large value of $\kappa_{\mathrm{SCE}}$ indicates a parameter pair with an extinction probability output that is very sensitive to input perturbations. 

Over this space, we highlight the maximum $\kappa_{\mathrm{SCE}}$ for a given $R_{0}$ (light blue open circle markers)  and a given $k$ (yellow open circle markers). For large values of $k$, the largest $\kappa_{\mathrm{SCE}}$ appear near $R_{0} \approx 1$. This matches conventional wisdom from the physics of phase transitions: The sensitivity of our forecasts aligns with the sensitivity of the dynamical system, and both are maximized at the critical point $R_{0} = 1$. Unexpectedly, for small values of $k$, particularly $ k<0.3$, the sensitivity of our forecasts is maximized at larger values of $R_0$, and forecasts of heterogeneous epidemic spread are more uncertain with larger values of $R_0$.

To validate the results shown in panel (a) of Fig. \ref{fig:negative_binom_cond}, we created three violin plots, displayed in panel (b), showing fluctuations around the mean forecast, or extinction probability, for three parameter pairs, $(R_{0} = 1.2, k = 8.05)$ (square marker), $(R_{0} = 1.4, k = 0.12)$ (triangle marker), and $(R_{0} = 3.8, k = 0.12)$ (diamond marker) with input distributions perturbed by very large noise $\delta = 2^{-2}$. This figure confirms that $\kappa_{\mathrm{SCE}}$ correctly increases as the fixed point becomes sensitive to perturbations. 

\begin{figure*}[th]
    \centering
    \includegraphics[width=0.95\textwidth]{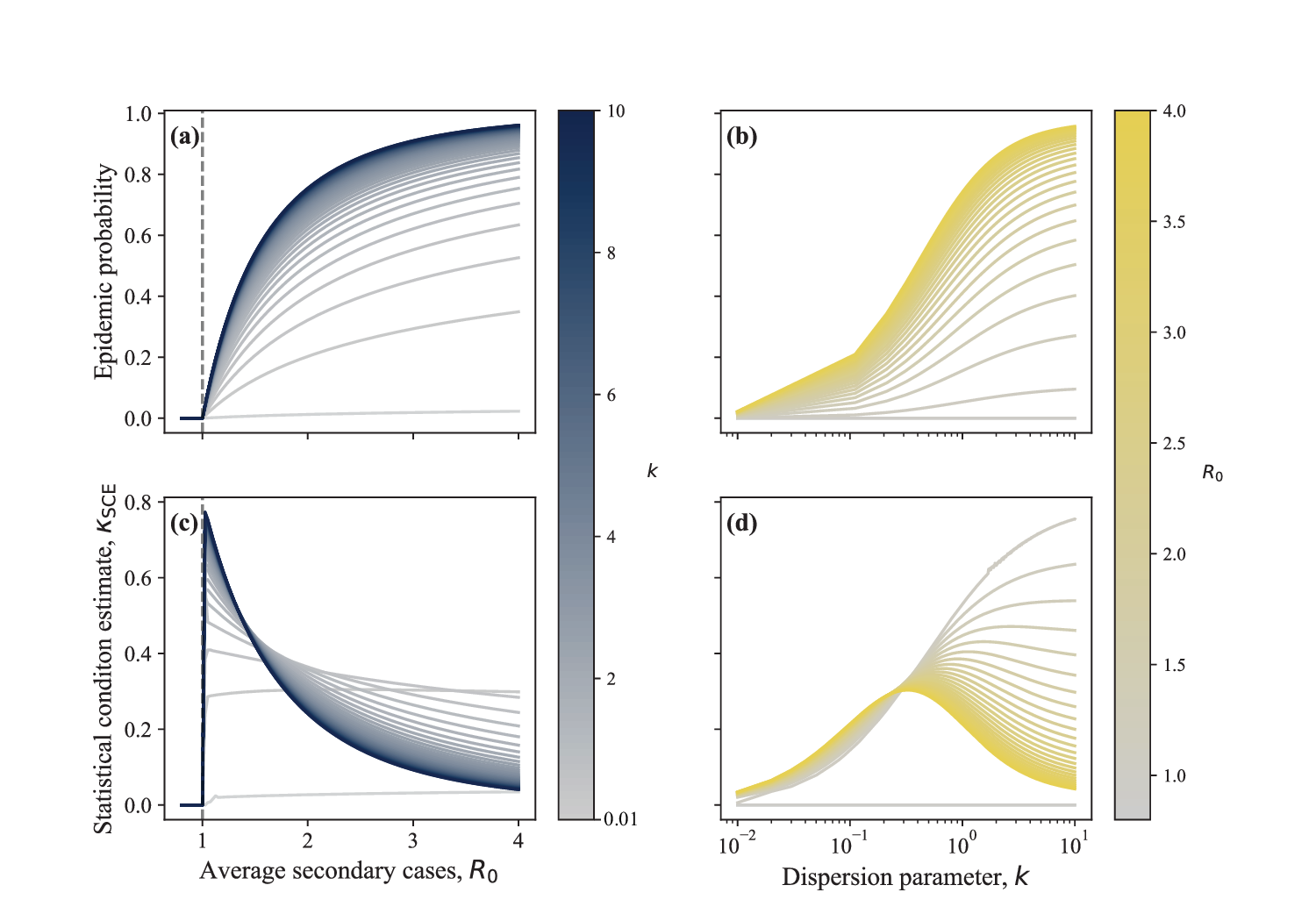}
    
    \caption[Negative binomial critical threshold and statistical condition estimate peaks:]{\textbf{Negative binomial critical threshold and statistical condition estimate peaks:} \textbf{(a)} Epidemic probabilities as a function of the average number of secondary cases, $R_{0}$, and \textbf{(b)} the dispersion parameter, $k$ when the other quantity is fixed. 
    The dotted line in panel (a) shows the critical transition threshold $R_{0} = 1$, above which a macroscopic outbreak happens regardless of the value of $k$. 
    The bottom panels illustrate how $\kappa_{\mathrm{SCE}}$ changes when we vary \textbf{(c)} $R_{0}$ and \textbf{(d)} $k$. The dotted line for $R_{0} = 1$ appears in panel (c) to showcase where the peak $\kappa_{\mathrm{SCE}}$ relates to the critical transition.  The plots in the bottom panels correspond to slices of Fig.~\ref{fig:negative_binom_cond}. 
    Panel (d) showcases how the change in $R_{0}$ over different $k$ values goes from a monotonically increasing curve to a non-monotonic curve.}
    \label{fig:negative_binom_peaks_plots}
\end{figure*}

Figure ~\ref{fig:negative_binom_peaks_plots} showcases slices of the parameter space, emphasizing the relationship between each individual parameter and the epidemic probability.  We again see that the largest $\kappa_{\mathrm{SCE}}$ appear near $R_{0} \approx 1$ for large values of $k$, in both panel (a) of Fig. \ref{fig:negative_binom_cond} and panel (c) of Fig. \ref{fig:negative_binom_peaks_plots}. Panels (a) and (c) of Fig. \ref{fig:negative_binom_peaks_plots} emphasize the conventional wisdom from phase transition theory, namely that peaks of sensitivity aligns with the epidemic threshold at $R_{0} \approx 1$. For $R_{0} > 1$ in Fig. \ref{fig:negative_binom_cond}, we notice a ridge of maximum $\kappa_{\mathrm{SCE}}$ develop as  $R_{0}$ increases. This ridge defines non-monotonic $\kappa_{\mathrm{SCE}}$ curves across values of $k$ when $R_{0} > 1$, which is clearly seen in panel (d) of Fig. \ref{fig:negative_binom_peaks_plots}. This maximum $\kappa_{\mathrm{SCE}}$ ridge for large $R_{0}$ values has associated values of $k$ that plateau at $k \approx 0.3$. 

We thus find $k \approx 0.3$ to be a pivotal threshold not only for the maximum $\kappa_{\mathrm{SCE}}$ for a given $R_{0}$, but also for the maximum $\kappa_{\mathrm{SCE}}$ for a given $k$. The latter maximum corresponds to $R_{0} \approx 1$ until $k \leq 0.3$. We notice the divergence of the maximum $\kappa_{\mathrm{SCE}}$ for a given $k$ at and below $k\approx 0.3$ in panel (a) of Fig. \ref{fig:negative_binom_cond}. This divergence is also displayed in panel (d) of Fig. \ref{fig:negative_binom_peaks_plots} as the curves associated with $R_{0} \approx 1$ dip lower than the curves of larger $R_{0}$ once $k \leq 0.3$. This contrasts the conventional wisdom previously mentioned, meaning the variation in the offspring degree distribution plays a vital role in the sensitivity of the epidemic forecast. 

Finally, to ground our results in examples, Fig. \ref{fig:negative_binom_cond} highlights three ranges of parameter values, shown by the dashed boxes. These parameter regimes correspond to the 2009 influenza outbreak in Italy, the 1958-1973 smallpox outbreak in Europe, and the 2020 SARS-CoV-2 outbreak in China. Table \ref{tab:diseases} details more information about the diseases listed above, along with another outbreak of SARS in 2003. Focusing on the influenza case first, the parameter regime sits in a heightened $\kappa_{\mathrm{SCE}}$ area, with values between 0.5215 and 0.5393, as noted in Table \ref{tab:diseases}. The precise parameter ranges for influenza imply that predictions based on the offspring distribution for influenza are quite sensitive to input perturbations, yet the sensitivity is consistent across the estimated parameter range. 

Unlike the influenza example, the SARS-CoV-2 and smallpox examples have larger parameter regimes, and are positioned at or below the $k\approx 0.3$ threshold. The smallpox regime encloses $\kappa_{\mathrm{SCE}}$ between 0.2571 and 0.3930.  Likewise, the span of the SARS-CoV-2 parameters also imply a large range of possible $\kappa_{\mathrm{SCE}}$. Even when assuming a specific estimate of $k \approx 0.1$, the $R_{0}$ range from the triangle marker,  $\kappa_{\mathrm{SCE}} = 0.1667$, to the diamond marker, $\kappa_{\mathrm{SCE}} = 0.2135$, shows approximately a 28\% increase of sensitivity when \emph{increasing} $R_{0}$. This reiterates our result that offspring distributions with $k \leq 0.3$ and larger $R_{0}$ are more sensitive than those with $R_{0} \approx 1$.

\begin{table*}[t]
    \renewcommand{\arraystretch}{1.3}
    \begin{tabular}{m{14em} | m{10em} | m{9em} | m{9em} | m{6em} | m{5em} }
     \hline
     \hline
     Disease - Location (Year) & Epidemic probability & $R_{0}$ (CI)$^{XX\%}$ & $k$ (CI)$^{XX\%}$ & $\kappa_{\mathrm{SCE}}$ & Ref. \\[0.5em]
     \hline 
     Smallpox - Europe (1958-1973) & 0.21-0.59 & 3.19 (1.66-4.62)$^{90\%}$ & 0.37 (0.26-0.69)$^{90\%}$ & 0.26-0.39 & \cite{lloyd2005superspreading}\\[1em]
     Influenza - Italy (2009) & 0.37-0.42 & 1.32 (1.30-1.34)$^{95\%}$ & 8.09 (5.17-11.79)$^{95\%}$ & 0.52-0.54 & \cite{dorigatti2012new}   \\[1em]
     SARS - Singapore (2003) & 0.00-0.46 & 1.63 (0.54-2.56)$^{90\%}$ & 0.16 (0.11-0.64)$^{90\%}$ & 0.00-0.45 & \cite{lloyd2005superspreading, leung2006seroprevalence, quah2004crisis}\\[1em]
      SARS-CoV-2 - China (2020) & 0.06-0.21 & 2.50 (1.4-3.9)$^{95\%}$ & 0.10 (0.04-0.2)$^{95\%}$ & 0.08-0.28 &   \cite{li2020early, endo2020estimating}\\
     \hline
     \hline
\end{tabular}
    \caption[Table of diseases defined by negative binomial parameters:]{\textbf{Table of diseases defined by negative binomial parameters:} This table provides specific cases of negative binomial parameters corresponding to disease outbreaks. Each instance details the corresponding epidemic probability and $\kappa_{\mathrm{SCE}}$ ranges. The table is inspired by Table 1 from \cite{hebert2020beyond} and Supplementary Table 1 from \cite{lloyd2005superspreading}. This table extends the analysis of H\'{e}bert-Dufresne \emph{et al.} with the $\kappa_{\mathrm{SCE}}$ for each disease. Epidemic probability ranges are calculated from the confidence interval boundaries. For a point of reference, the minimum $\kappa_{\mathrm{SCE}}$ value from this analysis is 0.00 and the maximum is 0.78.}
    \label{tab:diseases}
\end{table*}

% ~~~~~~~~~~~~~~~~~~~~~~~~~~~~~~~~~~~~~~~~~~~~~~~~~~~~~~~~~~~~~~~~~~~~~
\subsection{Case study: Transmission on random networks}
\label{sec:ERsims}
% ~~~~~~~~~~~~~~~~~~~~~~~~~~~~~~~~~~~~~~~~~~~~~~~~~~~~~~~~~~~~~~~~~~~~~

We next consider the SCE of epidemic models on heterogeneous contact networks as defined in Sec.~\ref{sec:pgfFormalism2}. We examine two ensembles of random networks characterized by their degree distributions. The first ensemble consists of ER random networks. In the infinite size limit, node degrees are independent random variables following a Poisson degree distribution, 
\begin{equation}
    p_{j} = \frac{\lambda^j e^{-\lambda} }{j!},\quad j\geq0\;,
\end{equation}
where $\lambda = \langle j \rangle$ is the mean degree. ER networks are therefore entirely defined by their mean degree  (varied between $\lambda \in [1,2]$ in this work), or equivalently, their density. They do not exhibit heterogeneity, hubs, correlations, or structural features beyond density. Since all moments of a Poisson distribution are equal, the degree distribution, generated by $G_0(x)$, and the excess degree distribution, generated by $G_1(x)$, are identical. Hence, the neighbors of any given node are statistically indistinguishable from any other node.   

Our second random network example, the scale-free network, is inspired by the fact that many real-world contact networks display extreme heterogeneity, where some nodes have many neighbors and most have few. This heterogeneity leads to heavy-tailed degree distributions and the formation of hub nodes within the network. Heterogeneous networks are often modeled using power-law degree distributions of the form,
\begin{equation}
    p_{j} = C j^{-\alpha},\quad j\geq1\;,
\end{equation}
where $C$ is a normalization constant. The exponent $\alpha$ controls the level of heterogeneity, and is varied between 2.1 and 3.6 in the present analysis. Larger values of $\alpha$ lead to more homogeneous networks, while smaller values result in networks with heavy-tailed degree distributions and the existence of hub nodes with disproportionately high degrees.  

\begin{figure*}[t!]
    \centering
    \includegraphics[width=\textwidth]{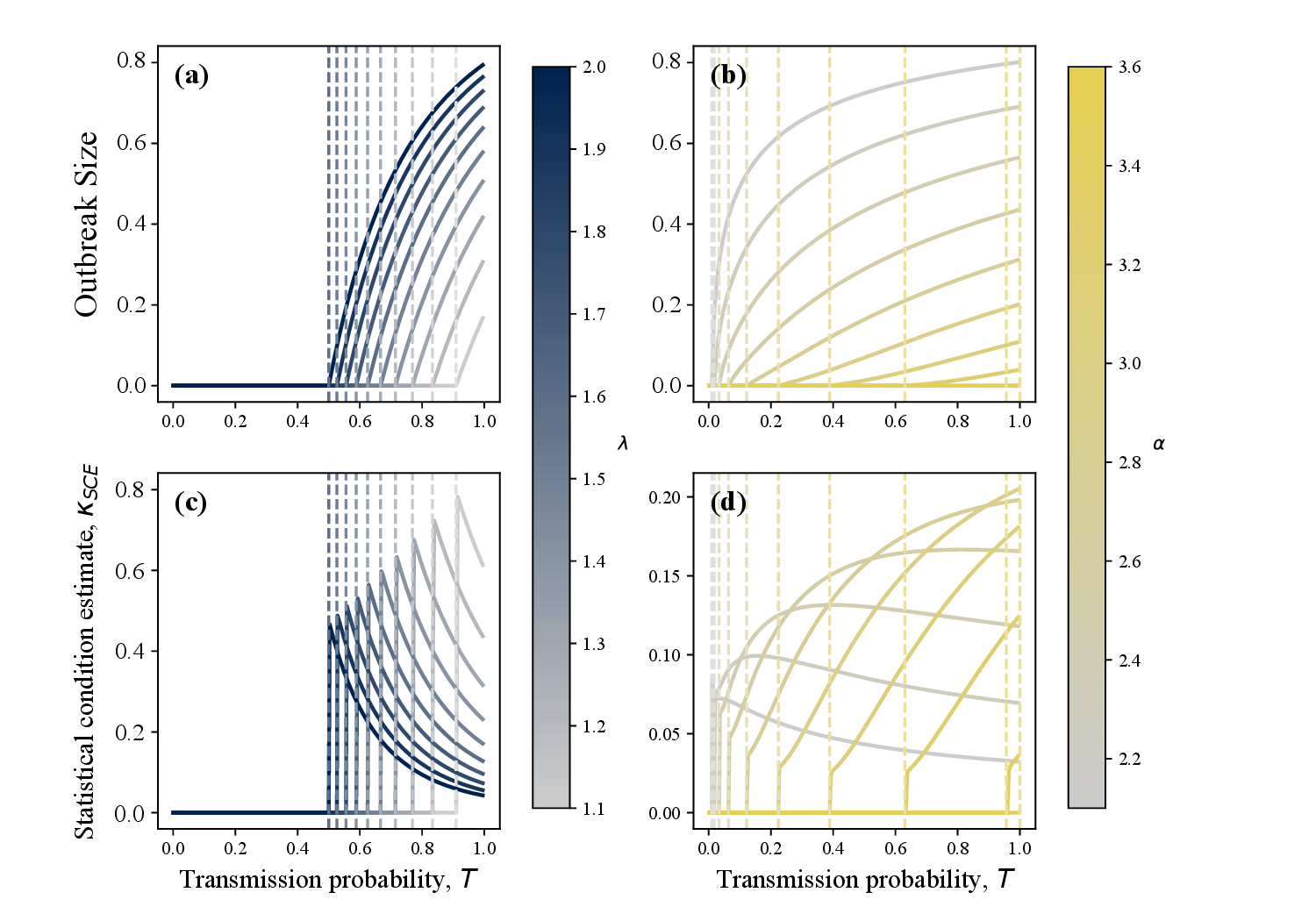}
\caption[Percolated random contact networks critical thresholds and statistical condition estimate peaks]{\textbf{Percolated random contact networks critical thresholds and statistical condition estimate peaks:}
 The outbreak size as a function of the infection rate $T \in [0,1]$ is displayed for \textbf{(a)} Erd\H{o}s–R\'{e}nyi (ER) networks and \textbf{(b)} power-law networks. The dotted lines indicate the critical transitions calculated using Eq. \eqref{eq:critcal_transition}. In panel (a), each solid line is associated with a value of the mean degree, $\lambda \in [1.1,2.0]$, which defines the network’s density. In panel (b), each solid line is associated with different values of the exponent, $\alpha \in [2.1,3.6]$, which defines the network’s heterogeneity. The $\kappa_{\mathrm{SCE}}$ for varying \textbf{(c)} $\lambda$ of an ER network are shown, and \textbf{(d)} for varying $\alpha$ of a power-law network are shown. The dotted lines for the critical transitions from panels (a) and (b) and displayed in panels (c) and (d) to relate the peak sensitivity to the critical transition thresholds.}
\label{fig:er_addative_condition_num}
\end{figure*}

The epidemic probability, $S = 1-u_p$, given in Eq. \ref{eq:rel_pandemic_size}, is determined by both the transmission probability, $T$, and the degree distribution parameter: Either the mean degree $\lambda$ for ER networks or the heterogeneity $\alpha$ of networks with power-law degree distributions. As in the case of branching processes, we use the method defined in Sec. \ref{sec:laub-xia} to obtain the $\kappa_{\mathrm{SCE}}$, or estimate of the sensitivity of the extinction probability $u_p$. ER and power-law networks display a critical transition in the epidemic probability, defined in Eq. \eqref{eq:critcal_transition}. Below a critical transmission probability, $T < T_c$, both network models have an epidemic probability of 0. And both network models have a nonzero epidemic probability above the critical transmission probability, $T > T_c$. 
As shown in Fig.~\ref{fig:er_addative_condition_num}, this critical transition marks a spike in the $\kappa_{\mathrm{SCE}}$. The figure displays the epidemic probability along with the $\kappa_{\mathrm{SCE}}$ for both ensembles as a function of the transmission probability and the respective parameter for the network degree distribution.

Similar to the results shown in Sec.~\ref{sec:negative_binom_lloyd-smith}, the largest $\kappa_{\mathrm{SCE}}$ for ER networks occur at the critical transition and decrease as the $T$ increases past $T_c$. As the mean degree $\lambda$ increases, the critical transition $T_c$ decreases. While the maximum $\kappa_{\mathrm{SCE}}$ always occurs at the critical transition, its value depends on the average degree $\lambda$. As the $\lambda$ increases, the $\kappa_{\mathrm{SCE}}$ at the critical transition decreases. Low-density networks have the highest peak and are thus the most sensitive to input perturbations in the degree sequence. 

Outbreak sizes and $\kappa_{\mathrm{SCE}}$ obtained on scale-free networks are shown for varying $T$ and varying $\alpha$, which, as a reminder, varies the heterogeneity. Panels (b) and (d) of Fig. \ref{fig:er_addative_condition_num} display results for scale-free random networks, where the behavior of percolation depends strongly on the value of $\alpha$. For $2 < \alpha < 3$, the first moment (average degree) is finite, but the second moment (average degree of neighbors) diverges, indicating extremely large fluctuations in node degree \cite{PhysRevLett.105.218701}. This results in a network with a few very large ``hub" nodes that dominate the spreading dynamics \cite{PhysRevLett.86.3200}. In the regime $\alpha < 3$, the critical transition is $T_c = 0$, which means that arbitrarily low transmission probabilities result in non-zero epidemic probabilities. The highest degree in a network of size $N$ scales faster than $\sqrt{N}$, and a small number of very large hubs dominate the epidemic dynamics \cite{PhysRevLett.105.218701}. 

If $\alpha > 3$, both the first and second moments of the degree distribution are finite. While hubs still exist, the degree distribution is less heavy-tailed, meaning that hubs do not dominate the epidemic dynamics to the same extent. The size of the highest-degree node scales more slowly than $\sqrt{N}$, and the largest hubs are not extreme enough to dominate the network. As a result, the critical transition is finite, $T_c > 0$ \cite{PhysRevLett.105.218701}. In dynamical systems, the moderately sized and weakly-coupled hubs in the $\alpha > 3$ regime ``trap'' outbreaks in their surrounding region in a phenomenon known as hub localization. Above the critical transition, increases in epidemic probability are driven by interconnections between microscopic hubs, making epidemic probability highly sensitive to perturbations in the degree sequence \cite{PhysRevLett.105.218701}. In percolation, a similar effect leads to smeared phase transitions, where system susceptibility can peak away from the critical point \cite{hebert2019smeared}.

This localization phenomenon, or smearing of the phase transition, explains the results in Fig. \ref{fig:er_addative_condition_num}. For $\alpha$ lower or close to 2, specifically $\alpha \lesssim 2.2$, the maximum $\kappa_{\mathrm{SCE}}$ occurs at the critical transition, $T_c = 0$, and decreases monotonically. For $\alpha \gtrsim 3$, the $\kappa_{\mathrm{SCE}}$ increases monotonically with $T$. In between, $2.2 \lesssim \alpha \lesssim 3$, $\kappa_{\mathrm{SCE}}$ has a first peak at the critical point and a second larger peak at a higher value of $T$. These results are remarkably different from those for ER networks, and the difference can be understood in analogy with the phenomenon of microscopic localization or smeared phase transition.

More generally, the results on scale-free networks mirror those of Sec.~\ref{sec:negative_binom_lloyd-smith}. In homogeneous systems, our mathematical forecasts for the probability of an epidemic are most sensitive at the epidemic threshold $R_0 = 1$ (e.g., in branching processes with $k>0.3$ or percolation on homogeneous networks or large enough coupling between hubs). In heterogeneous systems, however, sensitivity stems from a minority of superspreading events or hubs and can, therefore, be maximized with the higher average number of secondary cases, $R_0>1$.

% ~~~~~~~~~~~~~~~~~~~~~~~~~~~~~~~~~~~~~~~~~~~~~~~~~~~~~~~~~~~~~~~~~~~~~
\section{Discussion}
\label{sec:discussion}
% ~~~~~~~~~~~~~~~~~~~~~~~~~~~~~~~~~~~~~~~~~~~~~~~~~~~~~~~~~~~~~~~~~~~~~

Uncertain, noisy, or missing data all affect epidemiological models, obfuscate our interpretation of their forecasts, and hinder our interventions \cite{rosenblatt2020immunization}. When using PGFs to calculate the probability of an epidemic, it is necessary to quantify the sensitivity of our predictions to noise in model inputs. The results of this work answer this question for two epidemiological cases, one specifying a negative binomial branching process and the other specifying two random contact networks with bond percolation. 

For the negative binomial branching process model, we not only show that the forecasted probability of an epidemic is most sensitive to noise when $R_{0} \approx 1$ only if $k > 0.3$, but heightened sensitivity occurs for large $R_{0}$ when $k \leq 0.3$ as well.  As the dispersion parameter decreases, the offspring distribution reflects a higher potential for a superspreading event \cite{althouse2020superspreading}. Thus, when probabilities associated with a larger number of secondary infections are perturbed, the epidemic probability solution could be prone to more variation and be less stable. This result implies that our forecasts are most sensitive to noise not around the critical point between no epidemic and small epidemics, but at the distinction between small and large epidemics. This peak sensitivity can occur at very large $R_{0}$. Hence, public health officials must consider the sensitivity of heterogeneous offspring distributions even in the limit of a large basic reproduction number.

Switching to the bond percolation case study, the ER network example intuitively yields similar results to the negative binomial branching process case with $k > 0.3$. With these homogeneous networks, the largest sensitivity occurs around $T_c$ for varying $\lambda$. As $\lambda$ decreases, the peak sensitivity increases in conjunction with an increase in the corresponding $T_c$. 
Similar to the negative binomial branching process case, this presents a major obstacle in disease mitigation due to small $R_{0}$ producing small or zero change epidemics. On the other hand, the power-law random network analysis defines heterogeneous contact distributions or fat-tailed distributions. While there is a spike at the critical transition for all values of $\alpha$, the sensitivity continues to grow past each threshold for $\alpha \gtrsim 2.2$. Similar to the negative binomial branching process case, fat-tailed degree distributions create sensitive forecasts in outbreak size solutions, which is essential information for public health officials.

Heterogeneity of the degree distribution of a network, here parametrized by the scaling exponent $\alpha$, has well-documented impacts on the moments of the distribution and on the largest expected degrees \cite{PhysRevLett.105.218701}. Values $\alpha > 3$ lead to significant hubs with finite coupling such that epidemic dynamics can localize around the hubs \cite{stonge2018phase}, and the sensitivity of the model appears fundamentally different in that regime (as opposed to global, delocalized, epidemic spread). However, the mechanisms producing a similar behavior in the negative binomial branching process with $k\leq 0.3$ remain an open question. 

The conventional wisdom that the model's response to variations in input is maximized at the epidemic threshold is based on phase transition theory. The system moves from a disease-free state below the critical transition, $R_0 = 1$, to an epidemic state above it, and the response, or change in output, is therefore infinite at the threshold. Following this conventional wisdom, the quantitative forecasts of our models exhibit heightened sensitivity around that critical transition point where forecasts for the probability of an epidemic transition from exactly zero to barely above zero. However, we see a different picture in heterogeneous systems, i.e., in processes with superspreading events (dispersion parameter $k \leq 0.3$) or in transmissions on networks with fat-tailed degree distributions (scale exponents $\alpha \gtrsim 2.2$).
Interestingly, with enough heterogeneity, we find more sensitive forecasts at higher values of $R_0$ where small variations can change our prediction from a low to a high probability of an epidemic.
This occurs because changes in epidemic probability are relatively small at the critical point of heterogeneous systems, especially in systems with smeared phase transitions.
This does not mean that the conventional wisdom is completely wrong, but that system susceptibility is different from forecast sensitivity. These two concepts appear to align only in homogeneous systems.

\section*{Acknowledgments}
M.C.B. is supported as a Fellow of the National Science Foundation under NRT award DGE-1735316. W.H.W.T is supported by UVM's College of Engineering and Mathematical Sciences as well as National Institutes of Health Centers of Biomedical Research Administrative Supplement 2024 AWD 118 037649. C.M.D., J.-G.Y. and L.H.-D. acknowledge financial support from the National Institutes of Health P20GM125498 Centers of Biomedical Research Excellence Award.

\newpage
\bibliography{source}

\newpage
\appendix

\section{Solving Self-Consistent Equations}
\label{sec:solving_self_consitent_equations}

In this paper, the size of the outbreak or the probability of the epidemic is related to the solution $u$ of a self-consistent equation involving a PGF
\begin{equation}
    g(u) = u\;,  
\end{equation}
where $g(u)$ represents a generic PGF whose form depends on the details of the problem at hand. Solving for the smallest real positive root of the polynomial $f(x) = g(x) - x$ is equivalent to solving for a non-trivial solution to the self-consistent equation.
These self-consistent equations are often solved by direct iteration, since the PGF has only positive or zero coefficients. One simply starts with a random guess for the solution $u$ and iterates $u=g(u)$ until convergence to a certain tolerance threshold.
Another, often faster method is to use
Newton’s method, where the gradient of the function at a point is used to construct a first-order Taylor expansion, whose solution estimates the next iteration.

In this paper, however, we use Steffensen’s (or Aitken’s) method for root finding, as it does not require a gradient, yet achieves the same performance as Newton's method.
The iteration proceeds as,
\begin{equation}
    x_{n+1} = x_n - \frac{f(x_n)^2}{f\big(x_n + f(x_n)\big) - f(x_n)}\;.
\end{equation}

\section{Absolute and relative condition number}
\label{sec:cond_num_appendix}

This section reviews the condition number, its estimation, and the relationship between various definitions.

\subsection{Absolute condition number}

The \emph{absolute condition number} formalizes the notion of sensitivity for any differentiable function $f : \mathbb R^m \;\to\;\mathbb {R}$ at a point $\mathbf{p}$.
Concretely, it is defined by the norm of the gradient, since $\|\nabla f(\mathbf{p})\|_2$ is exactly the maximal rate of change of $f(\cdot)$ under an infinitesimal perturbation,
\begin{equation}
    \kappa_{\mathrm{abs}}(\mathbf{p}) = \|\nabla f(\mathbf{p})\|_2.
\end{equation}
This definition is sensible and can be used as is, but it can also be justified from first principles using the directional derivative, which measures the rate of change of $f(\cdot)$ along any unit direction $\zhat$.
We go through this derivation because it suggests algorithms that we use in the main text.

Consider a perturbation to $\mathbf{p}$ in an arbitrary direction, 
\begin{equation}
    \label{eq:app_p_tilde_def}
    \tilde{\mathbf{p}} = \mathbf{p} + \delta\,\zhat,
    \qquad
    \|\tilde{\mathbf{p}}-\mathbf{p}\|_2 = \delta.
\end{equation}
In this notation, the directional derivative of $f(\cdot)$ at $\mathbf{p}$ along $\zhat$ is
\begin{equation}
    \label{eq:dir-der-perturb}
    \nabla_{\zhat} f(\mathbf{p})
    = \lim_{\|\tilde{\mathbf{p}}-\mathbf{p}\|_2\to0}
      \frac{f(\tilde{\mathbf{p}}) - f(\mathbf{p})}{\|\tilde{\mathbf{p}}-\mathbf{p}\|_2}
    = \nabla f(\mathbf{p})^\top \zhat,
\end{equation}
and the gradient represents the direction of steepest ascent, given by $\max_{\zhat}\nabla_{\zhat}f(\mathbf{p}) =\nabla f(\mathbf{p}).$
Thus, computing the gradient norm can be reframed as finding the direction of steepest ascent and computing the norm of the corresponding directional derivative.

This perspective is useful when the gradient norm $\|\nabla f(\mathbf{p})\|_2$ is difficult to calculate, as is the case here.
(The function $f(\cdot)$ we are interested in is the solution of a self-consistent equation that is not easy to differentiate analytically.) 
The idea is to use the fact that any directional derivative carries information about the full gradient norm, as shown in Eq.~\eqref{eq:dir-der-perturb}.
For a random unit vector from the unit $m$-sphere, $\zhat\sim\mathrm{Unif}(S^{m-1})$, it is known that~\cite{kenney1994small}
\begin{equation}
    \label{eq:dir_div_estimator}
    \mathbb{E}\bigl[\,\|\nabla_{\zhat}f(\mathbf{p})\|_2\,\bigr] =  \omega_m\, \|\nabla f(\mathbf{p})\|_2.
\end{equation}
where $\omega_m$ is the Wallis factor defined in Eq. \eqref{eq:wallis_factor_def} of the main text.

Rearranging Eq. \eqref{eq:dir_div_estimator}, we obtain an estimator of the absolute condition number \cite{kenney1994small},
\begin{equation}
    \label{eq:absolute_SCE}
    \hat{\kappa}_{\mathrm{abs}} = 
    \frac{1}{\omega_m}\, \mathbb{E}\bigl[\,\|\nabla_{\zhat}f(\mathbf{p})\|_2\,\bigr].
\end{equation}
Evaluating this expression is now easy, as all the steps can be carried out numerically with sampling.
The algorithm proceeds as follows:
\begin{enumerate}
    \item[(i)] Create a $m\times r$ matrix of independently and identically distributed random variables, $\widetilde{Z}$, with entries drawn from a standard normal distribution $\mathrm{N}(0,1)$.
    \item[(ii)] Calculate the QR decomposition 
    \begin{equation*}
        \widetilde Z = Q\,R,
        \quad
        Q = [\,\mathbf z^{(1)},\ldots,\mathbf z^{(r)}\,],
        \quad
        Q^\top Q = I_{r}.
    \end{equation*}
    where $Q$ has dimensions $m\times r$ and $R$ has dimension $r\times r$. 
    \item[(iii)] Use the orthonormal columns $\{\mathbf z^{(i)}\}$ as perturbation directions.  For each $i$, compute
    \begin{equation*}
        \nu^{(i)}
        = \nabla f(\mathbf{p})\cdot\mathbf z^{(i)}
    \end{equation*}
    using Eq.~\eqref{eq:dir-der-perturb}, and estimate:
    \begin{equation}
    \label{eq:kappa_abs_est}
         \hat{\kappa}_{\mathrm{abs}} = \frac{\omega_r} {\omega_m} \sqrt{|\nu^{(1)}|^2 + \hdots + |\nu^{(r)}|^2}\;.
    \end{equation}
\end{enumerate}
This algorithm uses a QR decomposition to generate orthogonal samples, which we haven't discussed yet.
The idea is essentially to reduce the variance of $\hat\kappa_{\mathrm{abs}}$ to obtain a sharper approximation \cite{kenney1994small}. 
Since these perturbations form an orthogonal basis, each estimate $\nu_i$ represents the directional derivative along a specific basis vector. 
Equation~\eqref{eq:kappa_abs_est} thus represents the 2-norm of a gradient in a random subspace defined by the QR decomposition. 
As the dimension of this subspace approaches the dimension of the full $m$-dimensional domain of $f(\cdot)$, this approximation of the gradient approaches the true gradient.
The additional coefficient $\omega_r$ accounts for the fact that only a subspace is sampled when $r<m$.

In practice, one can use a fairly small value of $r$ to obtain good estimates.
Note that when computing the Wallis factor, we use the fourth-order expression \cite{kenney1994small}
\begin{equation}
    \label{eq:omega}
    \omega_{m} = \sqrt{\frac{2}{\pi(m - 0.5)}}\sqrt{\frac{184m^4}{1 + 23m + 23m^2 + 184m^4}}\;,
\end{equation}
since skip factorials grow rapidly with $m$, but their \emph{ratio} grows slowly enough to admit an accurate approximation.

\subsection{Relative condition number}
\label{sec:rel_cond_num_appendix}

The absolute condition number measures the absolute sensitivity to change of a function.
We need to make two modifications to adapt it to our problem.
First, it is often preferable to calculate the sensitivity of a function relative to the size of the perturbation---we want to capture the idea that small changes to the distribution of degrees or offspring can lead to large changes to the epidemic outcome.
Second, we need to account for the fact that $\mathbf{p}$ is in fact a probability vector, meaning that it must sum to one after it is perturbed.

To this end, we use the \emph{relative condition number}~\cite{laub2008statistical} by normalizing the numerator and the denominator 
\begin{multline}
    \label{eq:rel_condition_number}
    \kappa_{\textrm{rel}}(\mathbf{p}) = \\
    \max_{\mathbf{\tilde{p}}_{\textrm{rel}}} \left\|
    \lim_{\mathbf{\tilde{p}_{\textrm{rel}}} \to \mathbf{p}}  
        \frac{
            |f(\mathbf{\tilde{p}}_{\textrm{rel}}) - f(\mathbf{p})|
        }{
            |f(\mathbf{p})|
        } 
    \bigg/ 
        \frac{
            \|\tilde{\mathbf{p}}_{\textrm{rel}} - \mathbf{p}\|_2
        }{
            \|\mathbf{p}\|_2
        }
     \right\|_2,
\end{multline}
where we have expressed the gradient norm as a maximization of the norm of the directional derivative.

In this new equation, the perturbed vector $\mathbf{p}_{rel}$ is now normalized, and it is also a relative perturbation in the sense that larger entries are perturbed more strongly.
We express it in terms of the matrix $ \smat = \textrm{diag}(\mathbf{p}) $, a diagonal matrix with elements equal to the elements of the vector $\mathbf{p}$, such that the product $\smat \zhat $ is the element-wise product of $\mathbf{p}$ and $\zhat$, 
\begin{equation}
    \label{eq:q_vec}
    \mathbf{\smat} \zhat 
    = \textrm{diag}(\mathbf{p}) \cdot \zhat  
    = [p_1 z_1, p_2 z_2, \hdots, p_m z_m]^{T}\;,
\end{equation}
where the size of the perturbation in each direction of $z_{\ell}$ is scaled by the corresponding component of $p_{\ell}$. 
We then use $\mathbf{\smat \zhat}$ in place of $\zhat$ in Eq.~\eqref{eq:app_p_tilde_def}, and define the relative perturbation $\mathbf{\tilde{p}}_{\textrm{rel}}$, 
\begin{equation}
    \label{eq:relative_perturbation}
    \mathbf{\tilde{p}}_{\textrm{rel}} = \frac{\vecp + \delta \smat \zhat}{||\vecp + \delta \smat \zhat||_1}\;,
\end{equation}
where the denominator ensures that $\mathbf{\tilde{p}}_{\textrm{rel}}$ is a probability mass that sums to one.
(Some of the elements of $\mathbf{\tilde{p}}_{\textrm{rel}}$ can be negative in principle, but Appendix~\ref{sec:appendixC} shows that the calculation holds in the limit of large support, $m\gg 1$).

We can then use the same line of reasoning as before to arrive at a numerical estimate of this relative condition number.
This leads to
\begin{equation}
    \hat{\kappa}_{\mathrm{rel}}(\mathbf{p}) = \frac{1}{\omega_m} \mathbb{E}\left[\nu_{rel}(\mathbf{z})\right]
\end{equation}
where
\begin{equation}
    \nu_{rel}(\mathbf{z}) = \frac{\nabla f(\vecp) \cdot \smat \zhat}{f(\vecp)},
\end{equation}
and the expectation is taken over $\hat{\mathbf z}\sim\mathrm{Unif}(S^{m-1})$.
This expression can be evaluated by sampling as before.

\section{Asymptotic connection to standard SCE}
\label{sec:appendixC}

In our calculations, we consider condition numbers on probability distributions. 
Since we renormalize the probability distributions after a perturbation, it is important to show that the effect of the normalization is asymptotically small. 
Here, we demonstrate that it has little impact in the sense that
\begin{equation}
    \left\|\vecp + \delta \smat \zhat \right\|_1 \approx \left\|\vecp \right\|_1 = 1,
\end{equation}
in expectation.
To show this, we first write the 1-norm explicitly as
\begin{align}
    \left\|\vecp + \delta \smat \zhat \right\|_1 
    &= \sum_{k=0}^{m} \Big|p_k + \delta p_k z_k\Big|   \notag\\
    &= 1+ \delta \sum_{k=0}^{m} p_k z_k  \notag\\
    &= 1 + \delta  \mathbf{p} \cdot\zhat.
    \label{eq:norm_of_renorm}
\end{align}
If we can show that the last term is small, then we have our result.
This is exactly what Theorem 2.1 of Ref.~\cite{kenney1994small} does.
It establishes that
\begin{equation}
    \mathbb{E}[ |\mathbf{p}^\top \zhat|] = \omega_m \|\mathbf{p}\|_2.
\end{equation}
In our case $\mathbf{p}$ is a probability vector, and its 2-norm is bounded between $1/\sqrt{m}$ (uniform distribution) and $1$ (delta distribution), while $\omega_m$ scales as $1/\sqrt{m}$. 
Furthermore, we have taken $\delta$ to be a small constant.
Hence, we find that
\begin{equation}
   \mathbb{E}[  | \delta  \mathbf{p}^\top \zhat |] = \mathcal{O}\left(\frac{\delta}{\sqrt{m}} \right).
\end{equation}
In other words, the perturbation has a negligible impact on the norm of probability vector $\mathbf{p}$ in the limit of large supports ($m \to \infty)$.

% ~~~~~~~~~~~~~~~~~~~~~~~~~~~~~~~~~
\section{Algorithm}
\label{sec:algorithm}
% ~~~~~~~~~~~~~~~~~~~~~~~~~~~~~~~~~
The algorithm below summarizes the steps involved in calculating $\kappa_{\mathrm{SCE}}$.\\

\begin{algorithm}[H]
\caption{$\kappa_{\mathrm{SCE}}$ for the solutions to $u = G(u)$}
\label{alg:sce_roots}
\begin{algorithmic}
\Require PGF $G(x)$ generating the distribution $\{p_\ell\}_{\ell=0}^m$ representing the probability a random individual has $\ell$ neighbors or the number of offspring in a branching process.
\Ensure Statistical condition estimate $\kappa_{\mathrm{SCE}}$
\State Solve for solution $u^*$ to $u = G(u)$.

\State Generate the $m\times r$  matrix $\widetilde{Z}$ by drawing $\tilde{z}_{ij} \stackrel{iid}{\sim} \mathrm{N}(0,1)$ and orthonormalize this matrix to obtain $[\mathbf{z}^{(1)}, \dots, \mathbf{z}^{(r)}]$ via QR decomposition.
\State Choose $\delta$.
\For{$i = 1, \dots, r$}
    \State (a) Solve $\tilde{u}$ = $\tilde{G}(\tilde{u})$, where $\tilde{G}(u)$ has coefficients:
    \[
    \tilde{p}_\ell  = \frac{p'_\ell}{\sum p'_\ell} , \quad p'_\ell = p_\ell(1 + \delta \mathbf{z}_\ell^{(i)}).
    \]
    
    \State (b) Calculate the component-wise sensitivity:
    \[
    \nu^{(i)} = \frac{|\tilde{u} - u^*|}{\delta |u^*|}.
    \]
\EndFor
\State Compute the $r$-sample component-wise $\kappa_{\textrm{SCE}}$ vector:
\[
\kappa_{\mathrm{SCE}} = \frac{\omega_r}{\omega_m} \sqrt{|\nu^{(1)}|^2 + \dots + |\nu^{(r)}|^2}.
\]
\end{algorithmic}
\end{algorithm}

\end{document}